\documentclass[12pt]{article}
\usepackage{graphicx}

\begin{document}

\title{
\hfill {\rm\small MCTP-03-33} \\
\vskip-15pt~\\ The Phase of the Annual Modulation: Constraining the
WIMP Mass}

\author{Matthew J. Lewis and Katherine Freese\\ \\ {\rm\small {\it Michigan
Center for Theoretical Physics, University of Michigan, Ann Arbor, MI
48109, USA}}}

\date{}
\maketitle

\begin{abstract}
  
  The count rate of Weakly Interacting Massive Particle (WIMP) dark
  matter candidates in direct detection experiments experiences an
  annual modulation due to the Earth's motion around the Sun.  In the
  standard isothermal halo model, the signal peaks near June 2nd at
  high recoil energies; however, the signal experiences a phase
  reversal and peaks in December at low energy recoils.  We show that
  this phase reversal may be used to determine the WIMP mass.  If an
  annual modulation were observed with the usual phase ({\it i.e.},
  peaking on June 2nd) in the lowest accessible energy recoil bins of
  the DAMA, CDMS-II, CRESST-II, EDELWEISS-II, GENIUS-TF, ZEPLIN-II,
  XENON, or ZEPLIN-IV detectors, one could immediately place upper
  bounds on the WIMP mass of [103, 48, 6, 97, 10, 52, 29, 29] GeV,
  respectively.  In addition, detectors with adequate energy
  resolution and sufficiently low recoil energy thresholds may
  determine the crossover recoil energy at which the phase reverses,
  thereby obtaining an independent measurement of the WIMP mass.  We
  study the capabilities of various detectors, and find that
  CRESST-II, ZEPLIN-II, and GENIUS-TF should be able to observe the
  phase reversal in a few years of runtime, and can thus determine the
  mass of the WIMP if it is $\mathcal{O}(100$ GeV).  Xenon based
  detectors with 1000 kg (XENON and ZEPLIN-IV) and with energy recoil
  thresholds of a few keV require 25 kg-yr exposure, which will be
  readily attained in upcoming experiments.

\end{abstract}

\maketitle

\section{Introduction}

Extensive gravitational evidence suggests that a dominant fraction of
the matter in our Galaxy is nonluminous, or {\it dark}.  Although the
identity of this dark matter is currently unknown, it may consist of
Weakly Interacting Massive Particles (WIMPs).  Numerous experiments
aimed at the direct detection of these WIMPs are currently being
developed.  These experiments generically measure the nuclear recoil
energy deposited in a detector when an incident WIMP interacts with a
nucleus in the detector.  An important signature of halo dark matter
in direct-detection experiments is the annual modulation induced by
the Earth's motion with respect to the halo \cite{kt}.  In the
standard model of the dark halo, in which the velocity distribution of
the WIMPs is a Maxwellian distribution truncated at the escape speed
of the Galaxy, the modulation of the WIMP interaction rate at high
nuclear recoil energies is in phase with the motion of the Earth with
respect to the halo, peaking when the relative speed is maximal (June)
and reaching a minimum when the relative motion is minimal (December).
For low energy recoils, however, it is well known that that the
interaction rate experiences a phase reversal, peaking in December
\cite{hasenbalg, lewin, green, primack}.  This phase reversal occurs
below a particular {\it crossover recoil energy}, $Q_c$.  In this
paper we examine what information about the mass of the WIMP can be
obtained from this phase reversal.  We emphasize that this phase
reversal could constitute an important signature of a WIMP flux, and
we also point out that because the crossover recoil energy is a
function of the WIMP mass, the observation of the phase of annual
modulation immediately places an upper limit on the allowed WIMP mass.

Furthermore, if WIMP direct detection experiments with sufficient
energy resolution determine the crossover energy $Q_c$, the WIMP mass
could be independently measured using only observations of the phase
of the annual modulation.

This paper is organized as follows. In section 2, we briefly review
the basics of WIMP direct detection experiments and the isothermal
model of the dark halo.  In section 3, we explore the dependence of
the annual modulation on energy recoil, and revisit the phase reversal
that occurs at low recoil energies.  Finally, we demonstrate how this
observation may be exploited to place limits on the WIMP mass and to
obtain estimates of the mass directly.  We emphasize that observation
of this phase reversal will constitute an unambiguous signature of an
extraterrestrial WIMP flux.

%%%%%%%%%%%%%%%%%%%%%%%%%%%%%%%%%%%%%%%%%%%%%%%%%%%%%%%%%%%%%%%%%%%
\section{WIMP Direct Detection Experiments}
%%%%%%%%%%%%%%%%%%%%%%%%%%%%%%%%%%%%%%%%%%%%%%%%%%%%%%%%%%%%%%%%%%%
\label{sec:directdetection}

More than twenty collaborations worldwide are presently developing
detectors designed to search for WIMPs. Although the experiments
employ a variety of different methods, the basic idea underlying WIMP
direct detection is straightforward: the experiments seek to measure
the energy deposited when a WIMP interacts with a nucleus in the
detector \cite{goodmanwitten}.  If a WIMP of mass $m_\chi$ scatters
elastically from a nucleus of mass $m_N$, it will deposit a recoil
energy of $Q = (m_r^2v^2/m_N)(1-\cos\theta)$, where 

\begin{equation}
\label{eq:reduce}
m_r \equiv m_{\chi} m_{N}/ (m_{\chi} + m_{N})
\end{equation}

\noindent is the reduced mass, $v$ is the speed of the WIMP relative
to the nucleus, and $\theta$ is the scattering angle in the center of
mass frame. We may compute (following, {\it e.g.}, \cite{jkg,
gondologelmini, lesarcs}) the differential detection rate, per unit
detector mass ({\it i.e.}, counts/day/kg detector/keV recoil energy)
associated with this process,

\begin{equation}
\label{eq:detectrate}
\frac{dR}{dQ} = \frac{\sigma_0 \rho_h}{2 m_r^2 m_\chi} F^2(Q)T(Q,t)
\end{equation}

\noindent where $\rho_h = 0.3$ GeV/cm$^3$/c$^2$ is the halo WIMP
density, $\sigma_0$ is the total nucleus-WIMP interaction cross
section and $F(Q)$ is the nuclear form factor for the WIMP-nucleus
interaction that describes how the effective cross section varies with
WIMP-nucleus energy transfer.  We consider here only spin independent
interactions, wherein the target nucleus can be approximated as a
sphere of uniform density smoothed by a gaussian \cite{helm}, and the
resulting form factor is,

\begin{equation}
F(Q) = \frac{ 3 [ \sin(Qr_1) - Qr_1\cos(Qr_1) ] } {q^3 r_1^3} \,
e^{-Q^2s^2/2}
\end{equation}

\noindent where $r_1 = ( r^2 - 5 s^2)^{1/2}$, $ s \simeq 1$ fm, and,

\begin{equation}
r \simeq \bigl[0.91 (M/{\rm GeV})^{1/3} + 0.3 \bigr]
\times 10^{-13} {\rm cm}
\end{equation}

\noindent is the radius of the nucleus.  For spin-dependent
interactions, the form factor is somewhat different but again
$F(0)=1$.  In general, the form factor must be evaluated for each
detector nuclei.  For a more extensive discussion, see
\cite{goodmanwitten, kt, epv, ressell, jkg}.

For purely scalar interactions,
\begin{equation}
\label{eq:scalar}
\sigma_{0,\rm scalar} = {4 m_r^2 \over \pi} [Zf_p + (A-Z)f_n]^2
\,
.
\end{equation}
Here $Z$ is the number of protons, $A-Z$ is the number of
neutrons,
and $f_p$ and $f_n$ are the WIMP couplings to nucleons.
For purely spin-dependent interactions,
\begin{equation}
\sigma_{0,\rm spin} = (32/\pi) G_F^2 \mu^2 \Lambda^2 J(J+1) \, .
\end{equation}
Here $J$ is the total angular momentum of the nucleus and $\Lambda$ is
determined by the expectation value of the spin content of the nucleus (see
\cite{goodmanwitten,kt,epv,ressell,jkg}).

For the estimates necessary in this paper, we take the WIMP-nucleon
cross section\footnote{This is the cross section with nucleons at zero
  momentum transfer as discussed in Eq. (7.36) of \cite{jkg}.}
$\sigma_p = 7.2 \times 10^{-42}$ cm$^2$, and take the total
WIMP-nucleus cross section to be \footnote{In most instances, $f_n
  \sim f_p$ so that the following equation results from
  Eq.(\ref{eq:scalar}).}
\begin{equation}
\sigma_0 = \sigma_p \left(\frac{m_r}{m_{rp}}\right)^2 A^2
\end{equation}
where the $m_{rp}$ is the proton-WIMP reduced mass, and A is
the atomic mass of the target nucleus.

Information about halo structure is encoded into the quantity
$T(Q,t)$,

\begin{equation}
\label{eq:T(Q,t)}
T(Q,t) = \int_{v_{\rm min}} ^\infty \frac{f_d(v)}{v}dv,
\end{equation}

\noindent where $f_d(v)$ is the distribution of WIMP speeds relative
to the detector, and where $v_{\rm min} \equiv (Q m_N/2m_r^2)^{1/2}$
represents the minimum velocity that can result in a recoil energy Q.
To determine $T(Q,t)$, we must have a model for the velocity structure
of the halo.

%%%%%%%%%%%%%%%%%%%%%%%%%%%%%%%%%%%%%%%%%%%%%%%%%%%%%%%%%
\subsection{The isothermal halo} 
%%%%%%%%%%%%%%%%%%%%%%%%%%%%%%%%%%%%%%%%%%%%%%%%%%%%%%%%%

In what follows we will consider dark matter detection experiments
without directional capabilities; such experiments are sensitive only
to the WIMP flux integrated over the entire sky.  The most frequently
employed background velocity distribution is that of a simple
isothermal sphere.  In such a model, the galactic WIMP speeds with
respect to the halo obey a Maxwellian distribution with a velocity
dispersion $\sigma_h$,

\begin{equation}
\label{eq:halodist}
  f_h(v) dv = 4 \pi \left(\frac{3}{2\pi \sigma_h^2} \right)^{3/2} v^2
  \exp \left({-\frac{3v^2}{2\sigma_h^2}}\right)
\end{equation}

\noindent where $v$ is the WIMP velocity relative to the Galactic
halo, and where we have performed an angular integration over the
entire sky.  We take the velocity dispersion of our local halo to be
$\sigma_h = 270$ km/s.  This is the local velocity distribution; it is
expected to vary with spatial position throughout the Galaxy.

As pointed out by Drukier, Freese, and Spergel \cite{kt}, and studied
(in the context of an isothermal halo) by Freese, Frieman, and Gould
\cite{ffg}, Earth-bound observers of the dark halo will see a
different, time dependent velocity distribution as a result of the
relative motion of the Earth with respect to the Galaxy.  To take this
into account, we estimate the velocity of the Earth with respect to
the Galactic halo.  Neglecting the ellipticity of the Earth's orbit
and the non-uniform motion of the Sun in right ascension, we may write
the speed of the Earth with respect to the the dark WIMP halo as

\begin{equation}
\label{eq:veh}
v_{eh} = \sqrt{v_{es}^2 + v_{sh}^2 + 2v_{es}v_{sh} \cos(2\pi t -
\phi_h)}
\end{equation}

\noindent where the phase $\phi_h = 2.61 \pm 0.02$ corresponds to June
2nd $\pm$ 1.3 days (relative to $\phi=0$ on January 1).  The speed of
the Earth with respect to the Sun is $v_{es} = 29.8$ km/s, and the
speed of the Sun with respect to the halo is $v_{es} = 233$ km/s. More
precise expressions for the motion of the Earth and Sun may be found
in \cite{lewin, green}. Translating the distribution in Eq.
\ref{eq:halodist} into the distribution as seen by an earthbound
detector, $f_d(v_{eh})$, and integrating the resulting distribution in
Eq. \ref{eq:T(Q,t)}, one obtains

\begin{equation}
\label{eq:translate}
T_h(Q,t) = \frac{1}{2v_{eh}(t)}\left[{\rm erf}
      \left(\frac{\sqrt{2}(v_{\rm min} +
      v_{eh}(t))}{\sqrt{3}\sigma_h}\right) - {\rm
      erf}\left(\frac{\sqrt{2}(v_{\rm min} -
      v_{eh}(t))}{\sqrt{3}\sigma_h}\right)\right] .
\end{equation}

\noindent With $T_h(Q,t)$, one can now compute the recoil energy
spectrum associated with the isothermal halo, as well as the annual
modulation of that spectrum that results from the motion of the Earth
around the sun.  For the case of an isothermal halo, one generally
expects that, above a critical recoil energy, the annual modulation
will be in phase with the motion of the Earth with respect to the
halo, peaking near June 2nd.  In the following section, we will
demonstrate the dependence of the phase of the modulation on the
observed recoil energy.

%%%%%%%%%%%%%%%%%%%%%%%%%%%%%%%%%%%%%%%%%%%%%%%%%%%%%%%%%%%%%%%%%%%%
\section{The Phase Reversal at Low Recoil Energies}
%%%%%%%%%%%%%%%%%%%%%%%%%%%%%%%%%%%%%%%%%%%%%%%%%%%%%%%%%%%%%%%%%%%%
\label{sec:phasereversal}

\begin{figure}[tb]
\begin{center}
\includegraphics[scale=.65]{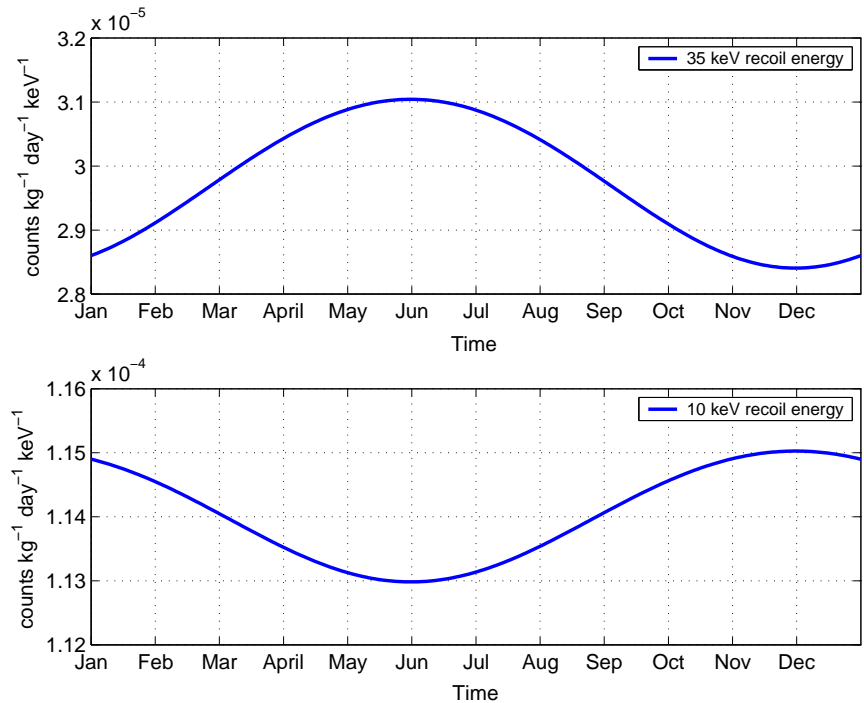}
\end{center}
\caption{The annual modulation of the WIMP differential detection rate
  for a WIMP with a mass of 65 GeV, in a $^{73}$Ge detector, for
  recoil energies of 35 and 10 keV.  Note the 180 degree phase shift
  between the two signals.}
\label{fig:diffevent}
\end{figure}

The annual modulation of the WIMP interaction rate \cite{kt, ffg}, and
in particular the phase reversal at low recoil energies
\cite{hasenbalg, green, lewin}, has been studied in detail by several
authors.  Here Fig. \ref{fig:diffevent} plots the expected
differential event rate for a WIMP of mass 65 GeV at recoil energies
of 35 keV and 10 keV.  We note that there is a 180 degree phase shift
between the two signals.  The origin of this effect may be understood
by considering a first order approximation to Eq. \ref{eq:translate}.
We note that the velocity of the Earth with respect to the WIMP halo,
$v_{eh}$, given in Eq. \ref{eq:veh} may be written, for $v_{es}\ll
v_{eh}$, as

\begin{eqnarray}
\nonumber v_{eh} &\sim& \eta_h + \epsilon_h(t).
\end{eqnarray}

\noindent where we define,

\begin{eqnarray}
\eta_h & = & (v_{es}^2 + v_{sh}^2)^{1/2}, \\ \epsilon_h(t) & = &
\frac{v_{es}v_{sh}}{(v_{es}^2 + v_{sh}^2)^{1/2}}\cos(2\pi t - \phi_h).
\end{eqnarray}  

\noindent Using this approximation,
%\begin{equation}
%\label{eq:translate25}
%T_h(Q,t) = \frac{1}{2(\eta+\epsilon(t))}\left[{\rm erf}
%      \left(\frac{\sqrt{2}}{\sqrt{3}}(\frac{v_{\rm min} + [\eta_h +
%      \epsilon_h(t)]}{\sigma_h})\right) - {\rm erf}
%      \left(\frac{\sqrt{2}}{\sqrt{3}}\frac{v_{\rm min} - [\eta_h +
%      \epsilon_h(t)] }{\sigma_h}\right)\right].
%\end{equation}
We may expand this expression about $\epsilon(t) = 0$, an
approximation that is well justified because $\epsilon_h(t)/\eta_h \ll
1$ for all $t$.  With the following definitions,

\begin{eqnarray}
\label{eq:xheh}
X_h^\pm &=& {\rm erf}\left(\frac{\sqrt{2}(v_{\rm min} \pm
\eta_h)}{\sqrt{3}\sigma_h}\right),\\ E_h^\pm &=&
\exp\left(-\frac{2(v_{\rm min} \pm \eta_h)^2}{3\sigma_h^2}\right),
\end{eqnarray}

\noindent we find that, to first order in $\epsilon$,

\begin{equation}
\label{eq:haloapprox}
T_h(Q,t) \approx \frac{1}{2}\left[\left(\frac{X_h^+ -
   X_h^-}{\eta_h}\right) + \epsilon_h(t)
   \left[\sqrt{\frac{8}{3\pi}}\frac{E_h^+ + E_h^-}{\sigma_h \eta_h} -
   \frac{X_h^+ - X_h^-}{\eta_h^2}\right]\right].
\end{equation}

\noindent We see that the differential detection rate will be
modulated by $\epsilon(t) \propto \cos(2\pi t - \phi_h)$, as expected.
We are particularly interested in the behavior of the amplitude of the
oscillation.  If we rewrite $T_h(Q,t)$ of Eq. \ref{eq:haloapprox} as

\begin{equation}
\label{eq:simpleform}
T(Q,t) = B + A \cos(2 \pi \omega t - \phi_h),
\end{equation}

\noindent we identify the amplitude of oscillation as

\begin{equation}
\label{eq:cofq}
A(Q)= \frac{v_{es}v_{sh}}{(v_{es}^2 +v_{sh}^2)^{1/2}}
  \left[\sqrt{\frac{8}{3\pi}}\frac{E_h^+ + E_h^-}{\sigma_h \eta_h} -
  \frac{X_h^+ - X_h^-}{\eta_h^2}\right].
\end{equation}

\begin{figure}[htb]
\begin{center}
\includegraphics[scale=.65]{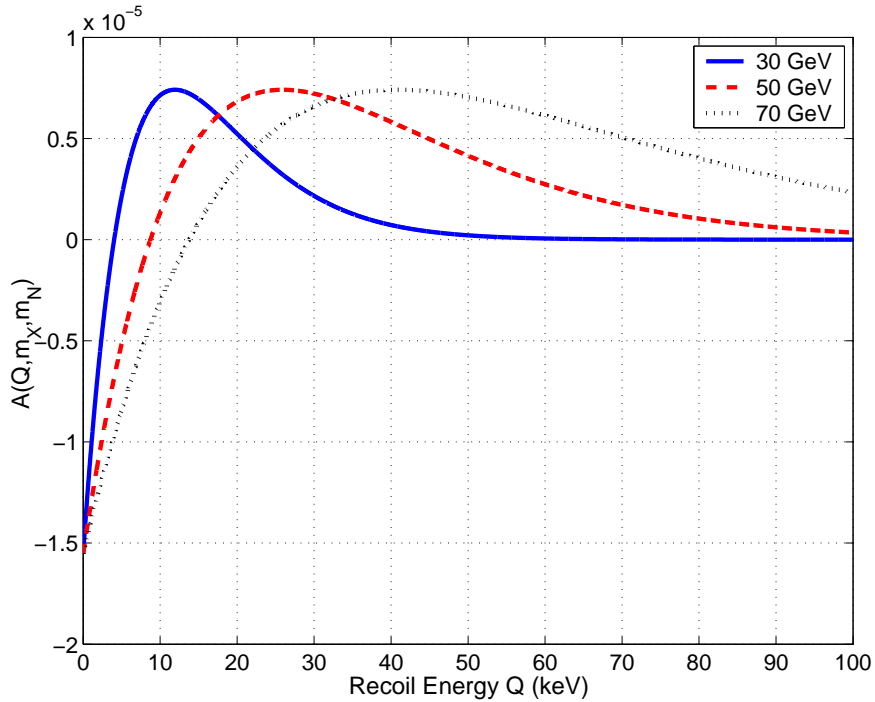}
\end{center}
\caption{The amplitude of modulation $A(Q)$ as a function of recoil
  energy $Q$ for WIMP masses of 30, 50, and 70 GeV in a $^{73}$Ge
  based detector.  For each WIMP mass there exists a particular recoil
  energy $Q_c$, at which the first order annual modulation of the WIMP
  differential event rate vanishes.  For $Q < Q_c$, we find that $A(Q
  < Q_c) < 0 $, and the differential WIMP interaction rate peaks in
  December, rather than June. }
\label{fig:coefficient}
\end{figure}

\noindent For $A(Q) > 0$, the modulation of the WIMP differential
event rate is in phase with the motion of the Earth with respect to
the halo, and will peak at June 2nd as usual.  For $A(Q) < 0$,
however, the modulation of the event rate moves 180 degrees out of
phase with terrestrial motion, peaking in December rather than June.
Values of $A(Q)$ are plotted in Fig. \ref{fig:coefficient} for $0 \le
Q \le 100$ keV, for a WIMP mass of 30, 50 and 70 GeV in a $^{73}$Ge
detector.  In general, for a WIMP mass $m_\chi$ and nuclear mass $m_N$
there will exist a nonzero, crossover recoil energy, $Q_c$, at which
$A(Q_c) = 0$, and below which $A(Q < Q_c) < 0$.  That is, {\it for $Q
< Q_c$ the WIMP interaction rate will peak in December rather than
June.}  We emphasize that these approximations, while illustrative,
give rise to errors of a few percent; in practice,
Eq. \ref{eq:translate} should be used for all computations.

The crossover recoil energy, $Q_c$ is a function of the mass of the
WIMP, as well as the mass of the target nuclei in the detector.  We
can estimate the critical recoil energy, below which we observe the
phase reversal. Fig. \ref{fig:crossover} plots the crossover recoil
energy as a function of WIMP mass, for NaI, $^{73}$Ge, $^{131}$Xe,
CaWO$_4$ and Al$_{2}$O$_3$ based WIMP direct detection experiments.

%%%%%%%%%%%%%%%%%%%%%%%%%%%%%%%%%%%%%%%%%%%%%%%%%%%%%%%%%%%%%%%%%%%%
\subsection{Placing an Upper Bound on the WIMP Mass}
%%%%%%%%%%%%%%%%%%%%%%%%%%%%%%%%%%%%%%%%%%%%%%%%%%%%%%%%%%%%%%%%%%%%

The observation of an annual modulation signal can be used to
constrain the WIMP mass.  For example, if a direct-detection
experiment observes an annually modulated signal peaking in June at an
energy recoil bin $Q_0$, we may place an upper limit on the WIMP mass
because we know that the crossover energy $Q_c$ must have been less
than $Q_0$, or a phase reversal would have been observed.  Because the
crossover energy is a monotonically increasing function of the WIMP
mass, this places an upper limit on the WIMP mass.

We will describe the upper bounds one can obtain in a variety of dark
matter detectors.  Of course, the experiments must have sufficient
exposure to see the annual modulation in order to obtain these bounds.

\begin{figure}[htb]
\begin{center}
\includegraphics[scale=.65]{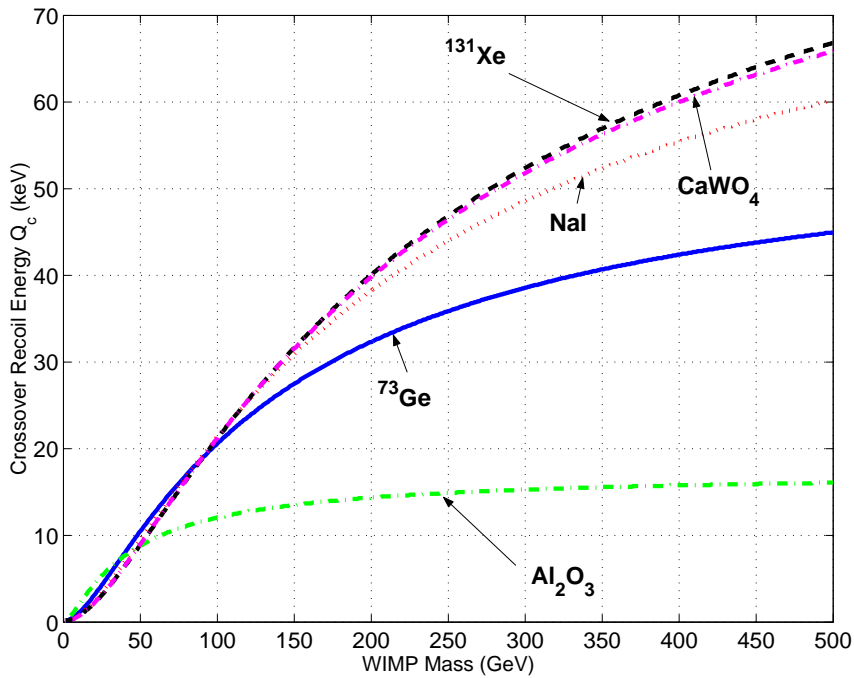}
\end{center}
\caption{The crossover recoil energy $Q_c$ as a function of WIMP mass
  $m_\chi$ for NaI, $^{73}$Ge, $^{131}$Xe, CaWO$_4$, and Al$_2$O$_3$
  detectors, assuming a halo velocity dispersion of 270 km/s.  The
  observation of an annual modulation signal peaking in June at a
  given recoil energy necessarily places an upper limit on the allowed
  WIMP mass. For example, the observation of a June-peaking annual
  modulation in the 22 keV energy recoil bin implies an upper limit on
  the WIMP mass of about 103 GeV for a NaI based experiment.}
\label{fig:crossover}
\end{figure}

As a concrete example, consider the DAMA experiment, which reports an
annual modulation in the event rate in its NaI detector in the 22-66
keV recoil energy bins \cite{DAMA}.  Although other experiments rule
out much of this region of parameter space \cite{EDELWEISS}, if one were
to believe the DAMA results, the fact that annual modulation in DAMA
peaks in June in this energy bin implies, from
Fig. \ref{fig:crossover}, that the maximum WIMP mass consistent with
the DAMA data is $m \leq 103$ GeV.

This same logic may be applied to other dark matter experimental
collaborations that may be sensitive to the phase reversal.  In
particular, in this paper we will consider the various versions of the
CRESST \cite{CRESST}, CDMS \cite{CDMS}, EDELWEISS \cite{EDELWEISS},
XENON \cite{XENON}, GENIUS \cite{GENIUS}, and ZEPLIN \cite{ZEPLIN}
experiments.  Table I collects the detector target materials, mass,
and recoil energy thresholds associated with these experiments.  In
Table II, we present the maximum WIMP mass consistent with an
observation of a June-peaking annual modulation in the lowest
accessible recoil energy bins of the respective experiments, as
determined using the arguments in the preceding paragraphs.

\begin{table}[tbp]
\label{tb:experiments}
\begin{tabular}{||c||c|c|c||}
\hline Collaboration & Material & $Q_{\rm thresh}$ (keV) & $M_D$ (kg) \\
\hline 
\hline

CDMS-II & $^{73}$Ge & 10\footnote{The quoted energy recoil threshold
of 10 keV is an ``analysis'' threshold imposed because events below 10
keV cannot be adequately discriminated.  The actual recoil energy
threshold of the experiment is 5 keV \cite{CDMS}.} & 5 \\

CRESST-II & CaWO$_4$ & $\sim 1$ & $10$ \\

DAMA & NaI & $22$ & 100 \\

EDELWEISS-II & $^{73}$Ge & $20$ & $38$ \\

GENIUS-TF & $^{73}$Ge & $1$ & $40$ \\

XENON & LXe & $4$ & $1000$ \\

ZEPLIN-II & LXe & $10$ & $40$\\

ZEPLIN-IV & LXe & $4$ & $1000$\\

\hline
\hline
\end{tabular}
\caption{Properties of various dark matter detection experiments
  including: the detector material, the detector energy recoil
  threshold, $Q_{\rm thresh}$, and the total detector mass, $M_D$. In
  the case of future experiments, the quantities are projected. }
\end{table}

\begin{table}[htbp]
\label{tb:MaxWIMPMass}
\begin{tabular}{||c||c||}
\hline Collaboration & Maximum allowed \\
& WIMP mass (GeV) \\

\hline 

CDMS-II & 48   \\

CRESST-II & 6  \\

DAMA & $103$ \\

EDELWEISS-II & 97 \\

GENIUS-TF & $10 $  \\

XENON & $29$ \\

ZEPLIN-II & $52$\\

ZEPLIN-IV & $29$\\

\hline
\hline
\end{tabular}
\caption{The maximum WIMP masses consistent with observing a
  June-peaking annual modulation signal in the lowest accessible
  energy recoil bins of the respective dark matter experiments. }
\end{table}

\vskip 1cm

%%%%%%%%%%%%%%%%%%%%%%%%%%%%%%%%%%%%%%%%%%%%%%%%%%%%%%%%%%%%%%%%%%%
\subsection{Detecting the Phase Reversal}
\label{sec:detectphase}
%%%%%%%%%%%%%%%%%%%%%%%%%%%%%%%%%%%%%%%%%%%%%%%%%%%%%%%%%%%%%%%%%%%

For WIMP direct detection experiments with sufficient energy
resolution and sufficient exposure, it may be possible to
determine the crossover recoil energy $Q_c$ itself, thereby making an
independent measurement of the WIMP mass.  If one determines the
recoil energy at which the modulation phase changes sign,
Fig. \ref{fig:crossover} may be used to estimate the WIMP mass.  

We now explore whether or not the phase reversal itself could be
practically observed in the context of current and near-future direct
detection experiments.  In order to extract a small annual signal from
the background, we require a large signal to noise ratio so that the
fluctuations are small compared to the desired signal.  This problem,
in the context of WIMP direct detection experiments, has been
discussed by Hasenbalg \cite{hasenbalg}.  Because the count rate for
direct detection experiments is so low, extended exposure times will
be required to reliably detect a small modulation.

We write the total signal as a function of time as,

\begin{eqnarray}
S(t) &=& \int_{Q_i}^{Q_f} \frac{dR}{dQ} dQ\\ &=& S_0(Q_i,Q_f) +
S_m(Q_i,Q_f) \cos(\omega t) + \mathcal{O}(S^2_m)
\end{eqnarray}

\noindent where $S_m$ is the amplitude of modulation, $S_0$ is the
unmodulated count rate, and $dR/dQ$ is defined as in
Eq. \ref{eq:detectrate}. These quantities will depend on the limits of
integration, $Q_i$ and $Q_f$, and may be be written in terms of the
differential event rate evaluated at its maximum in June, $S_J$, and
the rate at its minimum, in December, $S_D$,

\begin{eqnarray}
S_m = \frac{1}{2}[S_J - S_D] \hskip 1in S_0 = \frac{1}{2}[S_J + S_D].
\end{eqnarray}

If $S_m \ll S_0$, the theoretical signal to noise ratio may be written
as,

\begin{equation}
(s/n) \equiv \frac{S_m(Q_i,Q_f)}{\sqrt{S_0(Q_i,Q_f)}} \sqrt{MT}.
\end{equation}

\noindent where $M$ is the total detector mass and $T$ is the total
exposure time. As a reasonable criterion for distinguishing the
modulation signal from the noise, we 
 require that the $s$ be at least $2\sigma$ greater than the
statistical uncertainty.  This amounts to requiring $(s/n) = 2$.  For
low count rates, large $MT$ will be required to achieve this minimum
signal to noise ratio.

\begin{figure}[htb]
\begin{center}
\includegraphics[scale=.70]{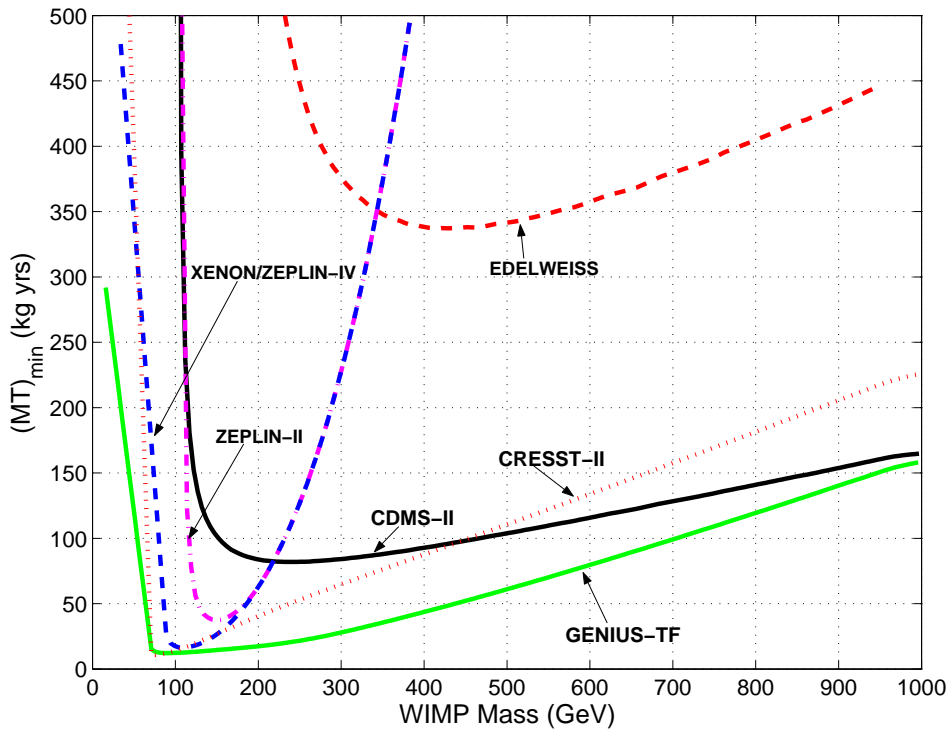}
\end{center}
\caption{The minimum exposure (in kg-yrs) required to observe the
  phase reversal in the WIMP-nucleus differential event rate at the
  $2\sigma$ level, as a function of WIMP mass for various detectors.}
\label{fig:s2n}
\end{figure}

In order to observe a phase reversal in a particular WIMP detector
with a given WIMP mass, an annual modulation peaking in December must
be detected in the energy range $Q_{\rm thresh}$ to $Q_c$, and a
modulation peaking in June must be observed in the energy range $Q_c$
to $Q_{\rm max}$, where $Q_{\rm thresh}$ is the recoil energy
threshold of the detector, and $Q_{\rm max}$ is the maximum recoil
energy accessible to the detector. In Fig. \ref{fig:s2n}, we plot the
minimum exposure $(MT)_{\rm min}$ (in kg-yrs) necessary to observe a
phase reversal for several current and future experiments, requiring a
minimum signal to noise ratio $(s/n) = 2$.  We should note that we are
assuming perfect energy detector resolution, and that these exposure
estimates therefore represent lower limits to the required exposure in
actual experiments.

We can determine the viablity of observing the phase reversal and
hence obtaining an estimate of the WIMP mass in various detectors.  We
note that, using this phase reversal technique, the experiments are
most sensitive to WIMP masses between 80 GeV to a few hundred GeV,
which is the most interesting mass range.  Accelerators (LEP-200)
place lower bounds on the neutralino mass of roughly 45 GeV.  The
upper limit for sensible supersymmetric dark matter particles is
roughly 1.5 TeV.  Recent upper bounds on the Constrained Minimal
Supersymmetric Model are tighter \cite{ellis2,rosz,edsjo}.  Hence the
most sensitive mass range in the detectors covers a compelling slice
of parameter space.

\subsection{Summary of Results}

As emphasized above, Fig. \ref{fig:s2n} presents the total exposure
time required to observe the phase reversal as a function of the WIMP
mass, for various detectors.

The CRESST-II experiment, which employs calcium tungstate (CaWO$_4$)
crystals, should be able to detect the phase reversal.  In particular,
for WIMP masses near 80 GeV, only two years of total exposure would be
required to observe the phase reversal.

For the $^{73}$Ge based detectors of CDMS and CDMS-II, the phase
reversal is not likely to be observed.  CDMS-II has projected
exposures of roughly 2,500 kg-days of data \cite{CryoArray}.  For the
most favorable WIMP masses, nearly 30,000 kg-days of data are required
to observe the phase reversal using germanium detectors with a recoil
energy threshold of 10 keV.

The EDELWEISS-II experiment, which also employs $^{73}$Ge, the
relatively high energy recoil threshold of the experiment renders it
unlikely to observe the phase reversal. At best, for WIMP masses near
400 GeV, the experiment requires $\sim 340$ kg-yrs of exposure.  With
its current mass, this will necessitate nearly 9 years of continuous
running time.

The GENIUS-TF experiment, a test facility for the projected GENIUS
experiment, is currently the best germanium-based platform for
observing the phase reversal \cite{GENIUS} .  Requiring only 10 kg-yrs
of exposure for WIMP masses $\sim 80-120$, the experiment could
observe the phase reversal with as little as a year of runtime. The
germanium based experiments have the advantage over other experiments
in that they are sensitive to a larger range of WIMP masses due to the
relative flatness of the $(MT)_{\rm min}$ curves in
Fig. \ref{fig:s2n}.  Other germanium based experiments not reflected
in the plot, such as the 0.2 kg HDMS experiment \cite{HDMS} do not
have sufficient exposure to observe the reversal.

The DAMA experiment, which is not represented in Fig. \ref{fig:s2n},
utilizes low mass NaI detectors with a high energy recoil threshold of
20 keV.  The DAMA collaboration will not be able to see the phase
reversal, despite its already extensive, 60,000 kg-day exposure. For
the most optimistic WIMP masses, DAMA would require nearly 1000 kg-yrs
of exposure to observe the WIMP phase reversal.

Liquid xenon experiments that are currently underway, such as the 40
kg ZEPLIN-II, will be sensitive to the phase reversal \cite{ZEPLIN}.
With a recoil energy threshold of 10 keV, the ZEPLIN-II experiment
will require only 40 kg-yrs of exposure for WIMP masses near 150 GeV.
For these WIMP masses, it may observe the phase reversal with only a
year of continuous running time.

The upcoming, large scale LXe detectors such as ZEPLIN-IV and XENON
with masses on the order of 1000 kg and energy recoil thresholds of a
few keV are in a excellent position to observe the phase reversal.
LXe detectors with energy recoil thresholds on the order of a few keV
require 25 kg-yrs exposure for WIMP masses in the range 100-150 GeV.
Hence, such experiments could observe a phase reversal with as little
as a year of continuous runtime.

As seen in Fig. \ref{fig:s2n}, a disadvantage of xenon based detectors
is that the minimum required exposure is low only for a narrow span of
WIMP masses. Outside of the 100-200 GeV mass range, the required
exposure rapidly becomes larger.  The other detectors, particularly
CRESST-II and GENIUS-TF, have flatter curves, {\it i.e.}, the minimum
required exposure does not change rapidly as a function of WIMP mass.
However, future 1 tonne xenon detectors, such as the XENON and
ZEPLIN-IV experiments, will have the advantage of sheer size, so that
they may reach the required exposures on reasonable timescales.

%%%%%%%%%%%%%%%%%%%%%%%%%%%%%%%%%%%%%%%%%%%%%%%%%%%%%%%%%%%%%%%%%%%
\section{Conclusions}
%%%%%%%%%%%%%%%%%%%%%%%%%%%%%%%%%%%%%%%%%%%%%%%%%%%%%%%%%%%%%%%%%%%

The reversal of phase at low recoil energy constitutes an unambiguous
signature of an incident WIMP flux.  The usual annual modulation,
while an important indicator of a WIMP halo, could conceivably be
mimicked by Earth bound physics (such as annually modulated
temperature fluctuations in the detector).  It is difficult, however,
to imagine non-WIMP scenarios that could engender a phase reversal
only at low energy recoils.  In addition, the critical recoil energy
depends in a well understood way on the WIMP mass, so that an
independent estimation of the WIMP mass can be obtained by locating
the critical recoil energy, $Q_c$, below which the phase reversal is
observed.  Furthermore, because the crossover recoil energy is a
monotonically increasing function of the WIMP mass, the observation of
an annual modulation in the WIMP signal will necessarily imply an
upper limit to the WIMP mass.

If a June-peaking annual modulation were observed in the lowest
accessible energy recoil bins of the DAMA, CDMS-II, CRESST-II,
EDELWEISS-II, GENIUS-TF, ZEPLIN-II or XENON/ZEPLIN-IV detectors, one
could immediately place upper bounds on the WIMP mass of [103, 48, 6,
97, 10, 52, 29] GeV, respectively, provided the experiments have
sufficient exposure to observe the annual modulation at all.

The observation of a phase reversal would provide independent
confirmation of the WIMP mass.  The CaWO$_4$ based detectors of
CRESST-II have the lowest required exposure, on the order of 10
kg-yrs, and would require only 2 years of running time to observe the
phase reversal.  With 40 kg of detector, GENIUS-TF and ZEPLIN-II could
observe the phase reversal with as little as a year of runtime.  We
note that the 1000 kg Xenon based detectors, ZEPLIN-IV and XENON,
operating with recoil energy thresholds on the order of a few keV
provide an excellent means of detecting the phase reversal in the
differential rate for WIMP masses on the order of 100 GeV, requiring
only 1 year of running time.  Hence, should the annual modulation be
observed, its phase reversal will provide a means to obtain the WIMP
mass in these detectors.

The analysis of this paper has assumed an isothermal halo model.
Galaxies, however, are thought to form hierarchically, with small
substructures condensing first and subsequently merging into
progressively larger structures.  It is therefore reasonable to think
that the dark halo of our galaxy may be populated with clumps or
streams of dark matter.  Indeed, work by Stiff, Widrow, and Frieman
\cite{stiff} suggests that there is a high ($\mathcal{O}(1)$)
probability that residual substructure exists at the solar radius,
with a contribution that is an additional few percent of the local
Halo density. In fact, as discussed by \cite{newberg1,newberg2}, the
Sagittarius dwarf galaxy, which is being tidally disrupted by the
Milky Way, is pouring dark matter down upon the solar neighborhood, at
the level of an additional (0.3-25)\% of the local Halo density.  The
peak of the annual modulation and value of the crossover recoil energy
will differ in halo models containing substantial clumps or streams of
dark matter, or characterized by anisotropic velocity distributions
\cite{forsco}.  In these alternate halo models, the phase and
amplitude of the annual modulation are generically altered, and the
impact on the crossover recoil energy must therefore be computed on a
case by case basis.  Such a study will be the subject of future
efforts.

We thank J. Frieman, H. Nelson, P. diStefano and L.  Stodolsky for
helpful remarks.  We acknowledge the DOE and the Michigan Center for
Theoretical Physics at the University of Michigan for support.

\end{document}